# *Evidence for an additive inhibitory component of contrast adaptation*

Kate S. Gaudry, PhD and Pamela Reinagel*, PhD.
*to whom correspondence should be addressed: preinagel@ucsd.edu; Section of Neurobiology, Division of Biological Sciences, University of California San Diego, La Jolla CA 92093.

## *Summary*

The latency of visual responses generally decreases as contrast increases. Recording in the lateral geniculate nucleus (LGN), we find that response latency increases with increasing contrast in ON cells for some visual stimuli. We propose that this surprising latency trend can be explained if ON cells rest further from threshold at higher contrasts. Indeed, while contrast changes caused a combination of multiplicative gain change and additive shift in LGN cells, the additive shift predominated in ON cells. Modeling results supported this theory: the ON cell latency trend was found when the distance-to-threshold shifted with contrast, but not when distance-to-threshold was fixed across contrasts. In the model, latency also increases as surround-to-center ratios increase, which has been shown to occur at higher contrasts. We propose that higher-contrast full-field stimuli can evoke more surround inhibition, shifting the potential further from spiking threshold and thereby increasing response latency.

## *Introduction*

Neurons in the early visual system must encode visual information across a broad range of temporal contrasts. Some neural properties are relatively conserved across contrasts. For example, the pattern of individual spiking events within a stimulus is similar across contrasts. Other neural properties have been reported to change with contrast. One such property is the response latency. Studies have reported that it takes longer for a visual neuron to respond to an excitatory stimulus during low contrast conditions (Baccus and Meister, 2002; Benardete and Kaplan, 1999; Chander and Chichilnisky, 2001; Kaplan and Benardete, 2001; Kim and Rieke, 2001; Levitt et al., 2001; Shapley and Victor, 1978; Solomon et al., 2004; Victor, 1987; Zaghloul et al., 2005). In 1978, Shapley et al. reported that the first-order response became less sharply tuned during lower-contrast stimuli and the phase of the response relative to a periodic stimulus was delayed. More recently, neural responses during different contrasts have been fit to Linear-Nonlinear cascades, and reports have confirmed that both the dynamics and the sensitivity of neurons depend on contrast (Baccus and Meister, 2002; Chander and Chichilnisky, 2001; Kaplan and Benardete, 2001; Kim and Rieke, 2001; Zaghloul et al., 2005). In retinal ganglion cells, the consistent finding is that responses have shorter latency and faster dynamics when stimulus contrast is high. The effects of contrast on response latency have not been examined in detail at the next stage of processing, in the LGN.

Additionally, it is known that the ON and OFF pathways are characterized by numerous physiological differences beyond a simply sign-inversion. For example, ON retinal ganglion cells can be characterized by a higher spontaneous firing rate, more linear light response, larger receptive fields, and faster kinetics than OFF retinal ganglion cells (Chichilnisky and Kalmar, 2002; Cleland et al., 1973; Kaplan et al., 1987; Passaglia et al., 2001; Troy and Robson, 1992; Zaghloul et al., 2003). Additionally, contrast has been reported to differentially affect the sensitivity of the two cell types in the retina. Further, while higher contrasts produced shorter latency responses for both cell types, this effect was more pronounced for OFF cells (Chander and Chichilnisky, 2001; Zaghloul et al., 2005). The interactions between the ON and OFF pathways provide further support for the differences between the pathways (Cohen, 1998; Renteria et al., 2006; Wassle et al., 1986; Zaghloul et al., 2003). Therefore, we predicted that contrast may differentially affect latencies of LGN cells within the two pathways.

In contrast to previous reports, we find that the response latency for cells of the LGN can decrease at lower contrasts. This surprising trend is particularly evident in ON cells. We propose that the paradoxical trend of shorter latencies at lower contrasts may be at least partially explained by a change in the distance-to-threshold with contrasts. A linear non-linear cascade analysis lends support to this hypothesis: at lower contrasts, spikes occur in response to a smaller generator potential (eg., excitatory stimulus strength) than do spikes at higher contrasts. Further support is offered by a realistic LGN spiking model that does not produce the unexpected latency effect when the threshold is fixed across contrasts but does when the distance to threshold is changed across contrasts. We show that our hypothesis can also predict the occurrence of low-contrast specific spiking events. Finally, Lesica et al. (2007) recently reported that the surround-to-center ratio of the spatial-temporal receptive fields can increase as contrast increases. We use the LGN spiking model to show that this change in the relative inhibition can predict the unexpected latency effects that we observed.

## *Results*

**Latency of ON cells Increases with Contrast**

We recorded from 41 relay cells of the LGN of anesthetized cats, while presenting full-field flickering binary white noise stimuli with each of three contrasts: 100%, 33% or 11% (see Methods). Responses from one representative OFF cell and one representative ON cell are shown in Figure 1. The times of the majority of spiking events are similar across contrasts, but we observe subtle changes in times of the spikes as contrast changes.

For the OFF cell, for most of the spiking events, the response latency increases as the contrast decreases, consistent with previous results. Interestingly, the ON cell's responses show the opposite trend: as contrast decreases, the response latency decreases.

In order to quantify the latency during each contrast condition, we estimate the linear filter of the cell as the spike-triggered average. We then define the latency as the time of the first peak of the spike-triggered average. In general, the response latency of ON cells increased as contrast increased, although some cell-by-cell variability was observed. In Figure 2A-F we show the spike-triggered averages during the high-, medium-, and low-contrast conditions for six ON LGN cells.

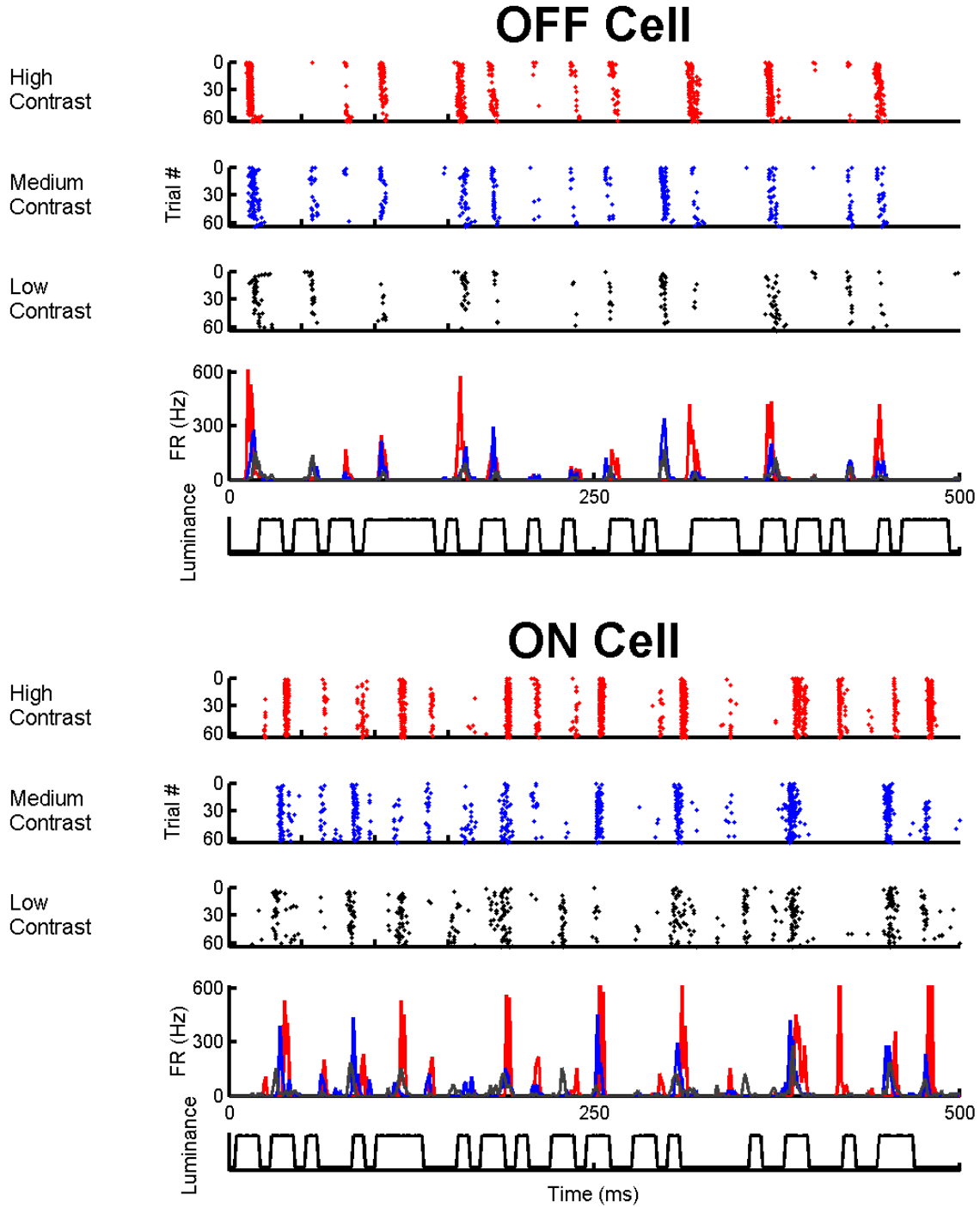

**Figure 1.** Responses of one ON and one OFF cell as a function of temporal contrast. One representative OFF Y LGN cell and one representative ON X LGN cell were stimulated by a full-field binary white noise visual stimuli of different contrasts. First row: Raster plot for 64 repeats of the same stimulus at 100% contrast. Each point represents the time of an action potential within that trial. A 150-ms segment is shown from the middle of the 5 s trials. Second and third rows: Responses to 64 repeats of the same full-field binary stimulus sequence, which was scaled about the mean to 33% and 11% contrast, respectively. Fourth row: Time varying firing rate in 1ms time bins during 100% (red curve), 33% (blue curve), and 11% (gray curve) contrast stimuli, derived from data in the raster plots. Bottom row: Binary stimulus presented at each of the three contrasts.

Figure 2A-B show two ON cells representing the most commonly observed latency trend amongst ON cells: the latency increased as contrast increased. That is, the time of the first peak of the spike-triggered average was later for high-contrast stimuli (red arrow) than for medium-contrast stimuli (blue arrow), and the time of the first peak of the spike-triggered average was later for medium-contrast stimuli (blue arrow) than for low-contrast stimuli (gray arrow).

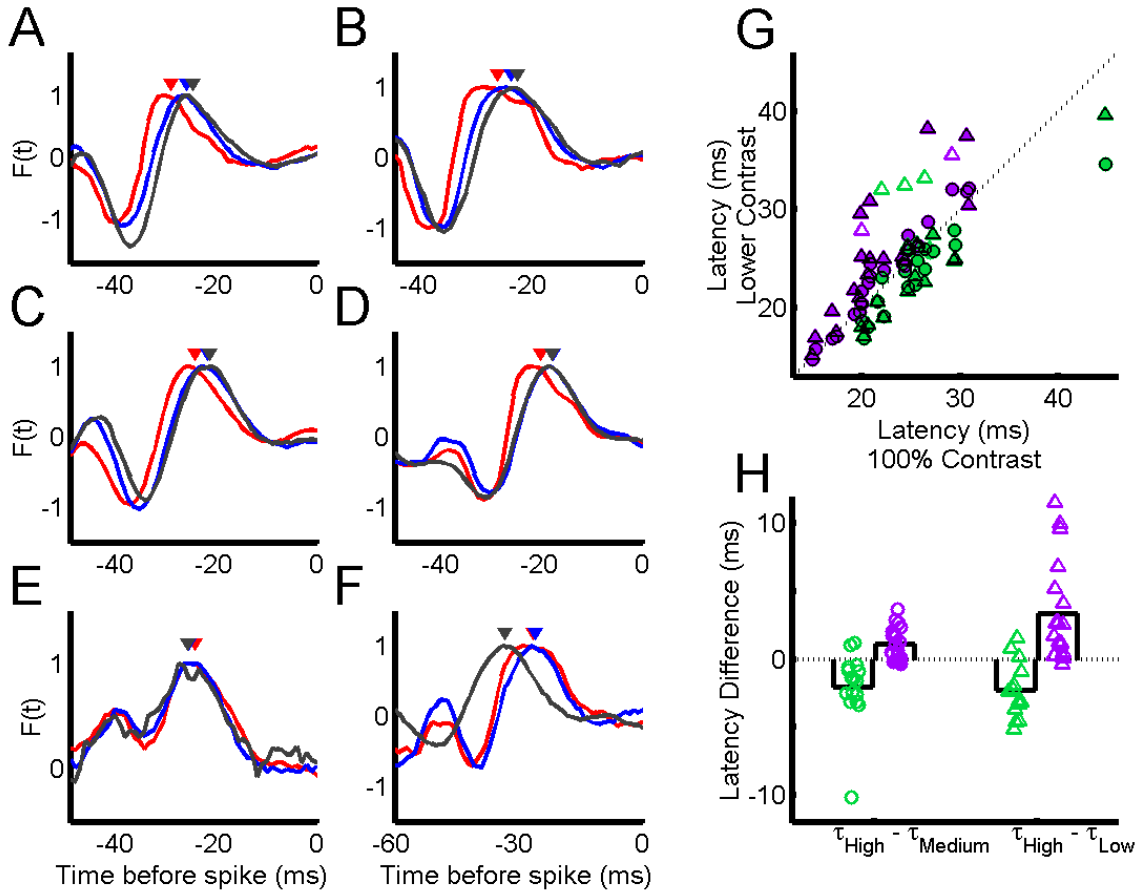

**Figure 2.** Effects of contrast on latency. A-F, The average stimulus preceding spikes, normalized by the amplitude of the first peak, from six representative ON LGN neurons calculated in the 100% contrast (red), 33% contrast (blue) and 11% contrast (gray) conditions. The arrows indicate the latency of the cell at each of the three contrasts. G, Response latency across cells. Latency of responses at 100% contrast (horizontal axis) is compared to latency of the same cell at either 33% contrast (○) or 11% (∆) contrast. ON and OFF cells are represented by green and purple symbols, respectively. Open symbols indicate cells for which – for at least one of the contrast conditions from the comparison – over 20% of the responses were bursts. Symbols above the diagonal indicate a latency decrease with contrast. H, Latency differences across cells. Shapes and colors of symbols are as described for panel G. Bars indicate the average latency difference across either the ON or OFF cells for the indicated contrast comparison. Cells for which over 20% of the responses were bursts (open symbols in panel G) are not shown.

For some ON cells, the latency increased as the contrast increased from medium contrast to high contrast (compare red and blue curves in Figure 2C-D) but showed minimal to no difference as the contrast increased from low contrast to medium contrast (compare blue and gray curves in Figure 2C-D). For a small fraction of the ON cells, contrast had only negligible effects on latency (Figure 2E). Finally, for another small fraction of the ON cells, the latency during low-contrast stimuli was longer than that during high-contrast or medium-contrast stimuli. We note that in these conditions, more than 20% of the low-contrast responses were bursts (see Discussion).

Across cells, we found that the effect of contrast on response latency depended on whether the neuron was an ON or OFF neuron. For OFF cells, the latencies from the high-contrast responses are significantly shorter than those from the medium- or low-contrast responses (purple symbols in Figure 2G-H; $p < 0.001$ for both comparisons). The decrease in latency from low to medium contrast was also significant ($p<0.01$). For ON cells, latency changed significantly in the other direction (green symbols in Figure 2C). The increase in latency from medium to high contrast was significant ($p<0.001$). While most cells also increased latency from low to high contrast, some cells that were characterized by high burst rates at low contrast showed the opposite trend, such that the increased latency trend was not significant ($p>0.05$).

**Predicted Contrast-Dependent Change in Resting Membrane Potential or Threshold**

In order to explore the latency results above, we propose that ON cells lie closer to their threshold at higher contrasts. We first consider a generator potential (convolved stimulus) at two contrasts. We convolve a segment of the high- and medium-contrast stimuli that were presented to the LGN cells with the spike-triggered average from the ON cell shown in Figure 1 to estimate the two generator potentials, $g_H(t)$ and $g_L(t)$. Figure 3A shows a segment of the generator potentials $g_H(t)$ and $g_L(t)$ corresponding to the same segment of the stimulus shown in Figure 1.

In a simple model, a cell fires a spike every time the generator potential $g(t)$ crosses a threshold. The dotted line in Figure 3A indicates an estimated threshold, which was calculated by identifying the threshold producing the highest correlation between the observed time-varying probability of spiking and convolved stimulus rectified at the threshold.

For high contrast, the fluctuations in the stimulus are large. Therefore, the generator potential $g_H(t)$ may quickly cross the threshold before, for example, reaching the peak of any given fluctuation. Meanwhile, for the lower contrast, the fluctuations are smaller, such that the generator potential $g_L(t)$ may need to wait until the most excitatory part of the fluctuation to reach threshold.

As explained above, the illustration of Figure 3A predicts that the generator potential $g(t)$ would cross the threshold later during low-contrast conditions than high-contrast conditions, resulting in longer latencies during low contrasts. While this is consistent with results reported for retinal ganglion cells, the opposite trend is observed for the ON LGN cells (Figure 2G-H).

Figure 3B illustrates the effect of reducing the distance-to-threshold for the lower-contrast condition. This can be accomplished either by adding a constant term to the generator potential $g_L(t)$ or by reducing the threshold for the lower-contrast condition. In this instance, we added a constant to the low-contrast generator potential $g_L(t)$ to obtain a modified generator potential $g_L'(t)$. (The constant was found by maximizing the correlation between the low-contrast time-varying firing rate and the modified generator potential $g_L'(t)$ rectified at the threshold). The same effect could be produced by increasing the distance-to-threshold for the higher-contrast condition.

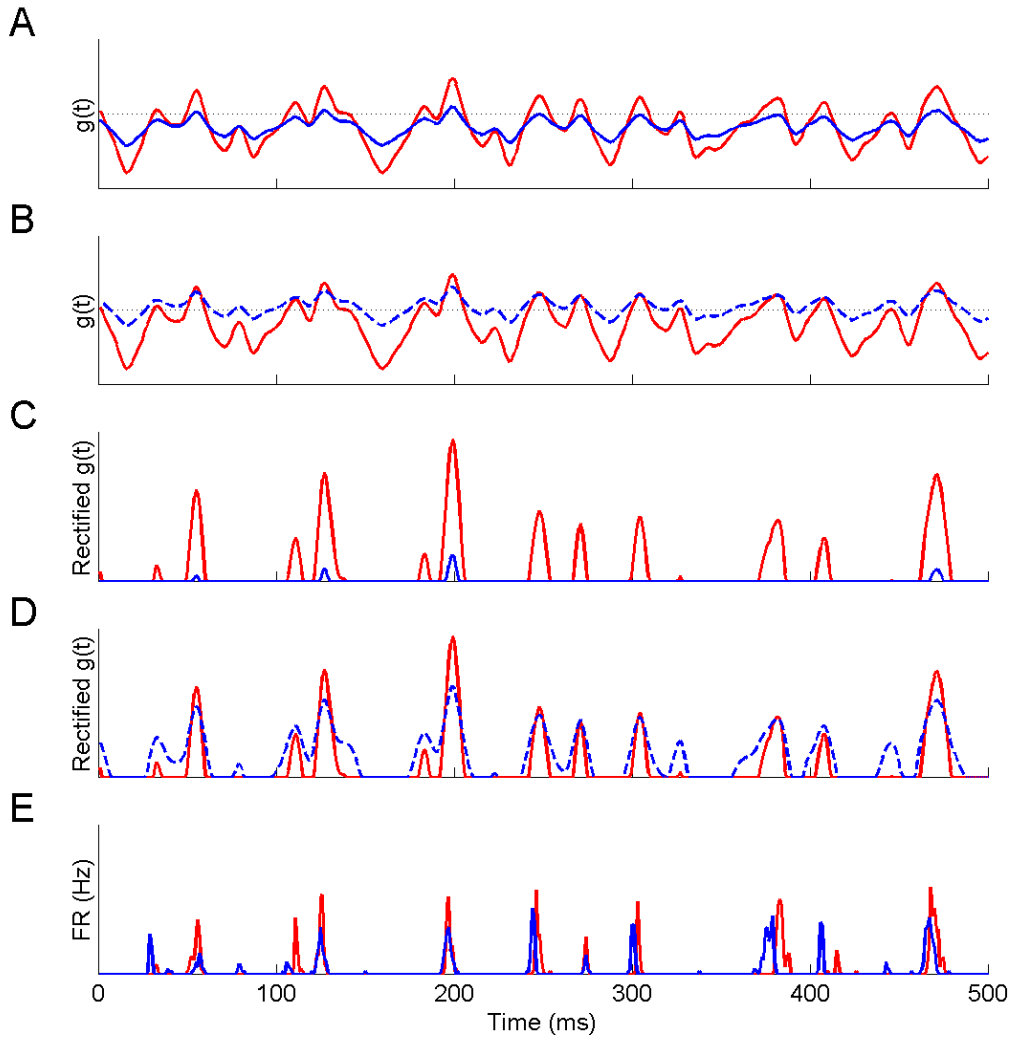

**Figure 3.** Effects of reducing distance-to-threshold of a generator potential. A, Generator potentials g(t) were calculated by convolving the stimulus at high- (red) or medium- (blue) contrast by the spike-triggered average of the ON cell from Figure 1B. The threshold (dotted line) was estimated as explained in Results. B, The modified generator potential (dotted line) was obtained by adding a constant value to the medium-contrast generator potential shown in A. C and D, the generator potentials shown in A and B are rectified at the threshold. E, The observed time-varying firing rate from the ON cell from Figure 1B for both contrast conditions.

The fluctuations in the modified generator potential $g_L'(t)$ remain smaller than those of the high-contrast generator potential $g_H(t)$. However, the reduction of the distance-to-threshold enables the modified generator potential $g_L'(t)$ to cross the threshold earlier than would the unmodified low-contrast generator potential $g_L(t)$. A sufficient reduction of the distance-to-threshold can result in the modified low-contrast generator potential $g_L'(t)$ crossing threshold before the corresponding high-contrast generator potential $g_H(t)$ would cross threshold.

We compared the rectified high-contrast generator potential and either the rectified unmodified low-contrast generator potential (Figure 3C) or to the rectified modified low-contrast generator potential (Figure 3D) to the observed probability of spiking (Figure 3E). Rectifying the generator potential $g(t)$ is consistent with a model that assumes the probability of spiking is zero until the generator potential $g(t)$ crosses a threshold, after which the probability of spiking is correlated with the generator potential $g(t)$. Notably, the rectified generator potential does not include refractory effects, which would truncate the spiking events and increase their reliability (Kara et al, 2000; Keat et al, 2001).

Subtle differences between the rectified generator potential and the probability of spiking are present, yet the rectified modified generator potential is able to produce similar latency changes observed in the neural responses of the LGN ON cell (blue peaks cross the threshold before the red peaks in Figure 3D). Meanwhile, the rectified unmodified generator potential cannot capture this latency trend (blue peaks cross the threshold after the red peaks in Figure 3C).

**Shift of Input-Output Functions Depends on Cell Type**

Above we speculate that the opposite latency effect observed in ON cells may be at least partially explainable by a decrease in the distance to threshold. We use a linear-nonlinear cascade to test this hypothesis in our LGN data. *See* Methods. Briefly, at each contrast, we estimate the linear filter of the neuron by calculating the normalized spike-triggered average (Figure 4A). The generator potential, $g(t)$, is defined as the stimulus convolved by this filter. By comparing the generator potential with the observed probability of firing across time bins, we estimate the Input-Output function (Figure 4B). The Input-Output function can be fit to a sigmoid. The amplitude of the sigmoid is related to the maximum probability of firing within a time bin. The slope of the sigmoid is related to the gain or sensitivity of the neuron: larger slopes indicate that the firing rate is more sensitive to stimulus fluctuations. The horizontal offset of the sigmoid can be thought of as being related to the difference between the baseline membrane potential and the threshold of the neuron: offsets further to the left indicate that a less excitatory stimulus will cause the neuron to spike.

Therefore, the hypothesized mechanism proposed to explain the latency result of the ON cells would also predict that the horizontal offset would change with contrast for the ON cells. The gain of neural responses is known to increase as stimulus contrast decreases (Benardete and Kaplan, 1997; Benardete et al., 1992; Brown and Masland, 2001; Chander and Chichilnisky, 2001; Jin et al., 2005; Kim and Rieke, 2001; Kremers et al., 2001; Shapley and Victor, 1978; Shou et al., 1996; Smirnakis et al., 1997; Solomon et al., 2004; Zaghloul et al., 2005) and we have previously reported the same trend for this data set (Gaudry and Reinagel, 2007a). Therefore, we do not hold either the gain or horizontal offset constant across contrasts, such that the observed gain changes will not confound horizontal offset changes. Still, we find that the Input-Output functions of the ON cell in Figure 4B differ by more than just the gain. This is demonstrated when we adjust the gain of the medium-contrast Input-Output function to match that at high contrast. The two nonlinear functions still differ in their horizontal offsets (Figure 4C and 4D).

We found that the effect of contrast on horizontal offsets depended on the cell type. For ON cells, the horizontal offset of the Input-Output function decreased as the contrast decreased (symbols below the diagonal in Figure 4E). A decrease in offset indicates that the input-output function shifted to the left, signifying that smaller generator potentials could evoke spikes at low contrasts than could at high contrasts. This shift is consistent with either a depolarization or a lower threshold at low contrasts. For OFF cells, the offset could increase, remain unchanged, or decrease (Figure 4F). Shifts in offsets for ON cells (average shift = -0.331 ± 0.324) differed significantly (p< 0.0001) from OFF cells (average shift = -0.0120 ± 0.242), while there was no significant difference between X and Y cells (p=0.38).

Many factors may interact to determine the effect that a change in the shift would have on latency. Nevertheless, we expected that cells characterized by large shifts in the offsets of the nonlinear functions would also show large changes in latency. Therefore, for each cell, we compared the difference in latency during two contrast conditions to the difference in offsets. (We excluded from this analysis data points in which over 20% of the responses were bursts, as bursts have different dynamics than tonic spikes). As shown in Figure 4G, the latency difference and offset difference were indeed correlated ($R^2$=0.15, p<.001). Cells whose latency was longer at higher contrasts (symbols below horizontal y=0 line) were also described by nonlinear functions that shifted to the left as contrast decreased (symbols to the left of vertical x=0 line).

In Figure 4I, we show that the observed differences in offsets can relate to substantially different time-varying firing rates. We compare predicted low-contrast time-varying firing rates using the linear-nonlinear model with various offsets to the observed low-contrast time-varying firing rate (Figure 4I; gray curve) from the ON cell from Figure 1 and from Figures 4A-D. For each of three conditions, we define the linear filter as the cell's low-contrast spike-triggered average. Further, we define the gain of the Input-Output function as the low-contrast gain calculated for the cell. We then separately calculate the predicted firing rate using the offsets calculated for that cell for each of the contrasts. Thus, differences in the predicted time-varying firing rates are entirely due to the contrast dependence of the offsets. The predicted time-varying firing rate associated with the low-contrast offset (black curve) provided a better match to the observed time-varying firing rate than either of the time-varying firing rates associated with the medium-contrast offset (blue curve) or the high-contrast offset (red curve). In this instance, the predicted time-varying firing rate associated with the high-contrast offset barely predicted any spikes at all.

Across cells, we calculated a set of first correlation coefficients between the predicted firing rates using the low-contrast offset and the observed low-contrast firing rate as well as a set of second correlation coefficients between the predicted firing rates using the high-contrast offset and the observed low-contrast firing rate. For OFF cells, the first and second sets of correlation coefficients were not significantly different from each other (p=0.34). For ON cells, the first set of correlation coefficients was significantly higher than the second set of correlation coefficients (p<0.01). Thus, for ON cells, contrast-dependent shifts in the offsets of the input-output functions can contribute to significant differences in time-varying firing rates.

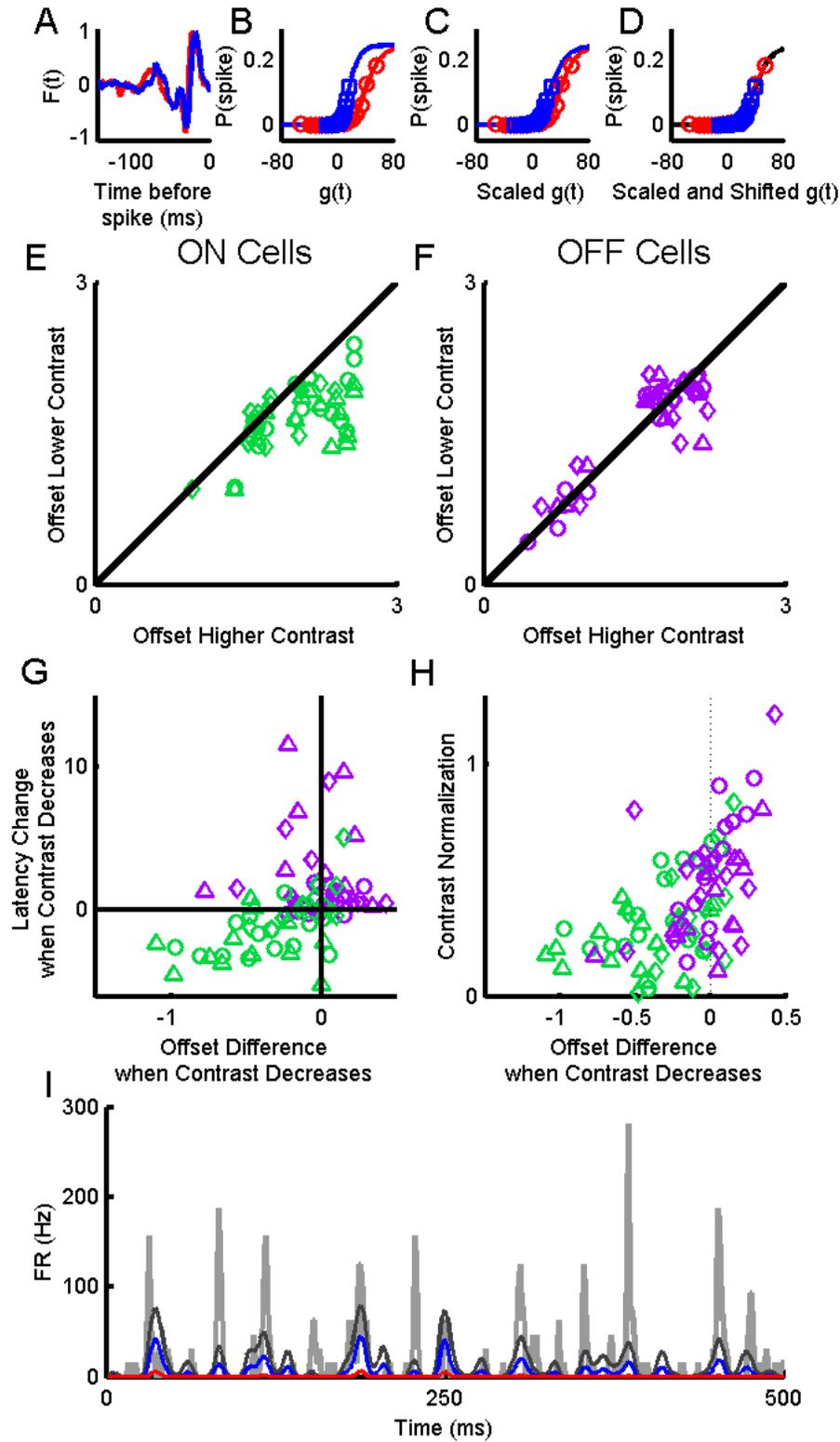

**Figure 4.** Shift in Input-Output functions with contrast. LGN responses were fit to a Linear-Nonlinear cascade at each contrast. Panels A-D illustrate the method for one Y OFF cell. A, Linear filters from a Linear-Nonlinear cascade model, fit to high- (red curve) and medium- (blue curve) contrast data. The filters were normalized by the amplitude of their first peak. B, Symbols indicate the observed probability of spiking versus the generator potential, *g(t)*, for the same cell as in (A). The curves show the best-fit sigmoid functions. Based on these fits, the gain at the high contrast was 0.63 times that at medium contrast. C, Equivalently, we can scale the generator potential at medium contrast by a factor of 0.63 in order to set the gain of the two input-output functions to the value at high contrast. For this cell, the nonlinear functions also differed in their horizontal offsets. D, After shifting the blue curve horizontally to correct for the difference in the horizontal offset (which differed by 0.97 units), the same sigmoid function (black curve) can now describe both the medium contrast's shifted and scaled nonlinear function ($R^2 = 0.996$) and the nonlinear function at high contrast (red symbols; $R^2 = 0.991$). E, Across all ON cells, we compare the offset between 100% and 33% contrast (○), 33% and 11% contrast (◊), and 100% and 11% (∆) contrast, where the gain at the higher contrast is always shown on the horizontal axis. If all differences in the nonlinearity were captured by changes in gain, offsets would be constant across contrasts (x-y, thick, solid line). Symbols below the line indicate the nonlinear function was shifted to the left at lower contrasts. F, We compare changes in offset with contrast for all OFF cells. Symbols as in (E). G, Difference in latency, shown in Figure 2C, compared to the difference in offsets from the nonlinear functions (shown in Figure 4E and 4F). Both differences are calculated as the variable at the lower contrast minus the variable at the higher contrast. Therefore, symbols above the horizontal x=0 line indicate that latency increased as contrast decreased, and symbols to the left of the vertical y=0 line indicate that the nonlinear function at the lower contrast was shifted to the left of that at the higher contrast. Each symbol represents a comparison between two contrast conditions for one cell. Symbol shapes indicate whether the comparison was between 100% and 33% contrast (○), 33% and 11% contrast (◊), or 100% and 11% (∆). Symbol colors are as defined in (C). H, Anti-correlation of multiplicative and additive gain changes. Symbols as in (G). I, Time varying firing rate (light gray line) in 1ms time bins of the ON cell from Figure 1 and panels A-D during a low-contrast stimulus. Colored lines indicate firing rates predicted by the linear-nonlinear model using offsets from the high (red curve), medium (blue curve) or low (dark gray curve) contrasts.

## Anti-Correlation of Multiplicative and Additive Gain Change

It is well known that the gain of an LGN neuron can change as the contrast changes, such that the neuron is more sensitive to smaller fluctuations at lower contrasts. Typically, gain changes are referred to as a multiplicative quality: it is thought that the neuron essentially amplifies or multiplies a signal at low contrasts. This differs from the additive gain change illustrated in Figure 3. Only the additive gain change can account for the latency trends reported herein. However, we suggest that a combination of the multiplicative and additive gain changes can occur.

We hypothesize that cells that exhibit less contrast normalization (multiplicative gain change) may exhibit a larger change in resting membrane potential or threshold (additive gain change). As described above, these properties can be separately measured for each neuron using the linear-nonlinear model. *See* Figure 4A-D.

We calculated the extent of contrast normalization κ by comparing the change in the neuron's sensitivity or gain to the change in contrast. If the gain increased by the same factor that the contrast decreased, the contrast normalization κ is defined as 1, while if the gain did not change as contrast changed, the contrast normalization κ is defined as 0. *See* Methods.

We find a significant correlation between the contrast normalization and the offset difference of the nonlinear functions described above (Figure 4H; $R^2 = 0.30$, p<0.0001). Some neurons show little contrast normalization and large offset differences (symbols in bottom left corner of Figure 4H), while others show large contrast normalization and little offset differences (symbols in top right corner of Figure 4H). Still the majority of neurons are characterized by a combination of the multiplicative and additive mechanisms.

**Latency Changes in LGN Spiking Model**

Above, we provide an illustration to show how changes in the generator potential may produce the opposite latency effects that we observe in ON and OFF cells in our experimental data (see Figure 3 and related Results section). However, as mentioned, the convolved stimulus does not incorporate other factors known to influence spiking responses, such as refractoriness and noise. Because ON cells' firing rate increases with contrast more than OFF cells, it was possible that longer-latency responses at high contrasts are due to the more frequently induced absolute or relative refractory periods.

Therefore, we implemented a nonlinear model that has been shown to successfully reproduce responses of LGN neurons to white noise stimuli (Gaudry and Reinagel, 2007b; Keat et al., 2001). This model includes both noise terms and refractory terms (see Methods). A large population of model cells (N=550) was created by varying the model parameters (see Methods). The same high- and medium-contrast stimuli that were presented to LGN neurons were used to generate model cell responses.

In the first condition, all parameters remained fixed across the contrasts for each model cell. We have previously shown that even with fixed parameters, the model replicates the multiplicative contrast normalization associated with contrast gain control (Gaudry and Reinagel, 2007a, 2007b). The latency was determined for each model cell at each contrast as described for Figure 2, by measuring the time of the first peak of the spike-triggered average. As shown in the contour density plot of Figure 5A and in the dashed curve in Figure 5C, the only contrast-dependent latency changes observed across the population of model cells were decreases in latency during higher contrast stimuli (the distribution indicated by the dashed curve is strictly to the left of 0). This suggests that noise and refractoriness are not sufficient to explain the opposing latency increase observed in LGN ON cells. In the second simulation condition, the threshold was lowered during the lower-contrast condition. We note that in the model this is formally equivalent to depolarizing the resting potential relative to a fixed threshold. The change in threshold was sufficient to produce both the increasing and decreasing latency trends with contrast, as shown in Figure 5B and the solid curve in Figure 5C.

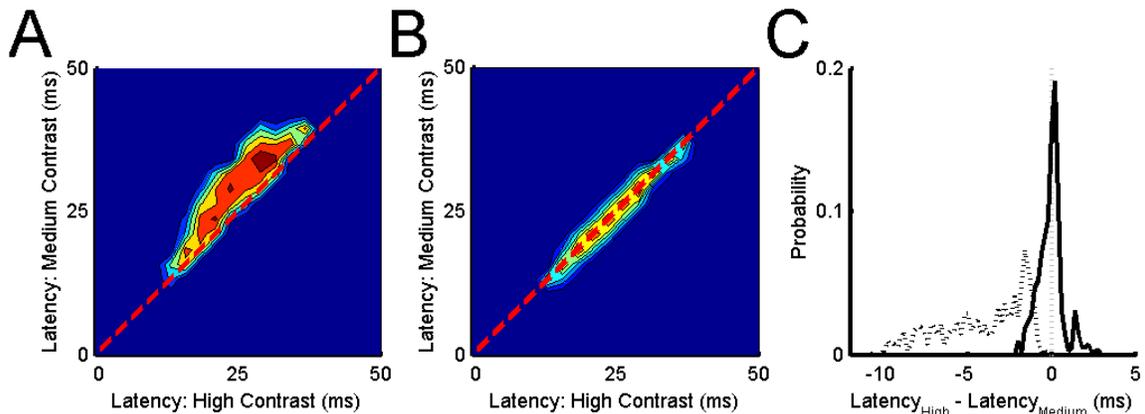

**Figure 5.** Latency trends in a spiking LGN model. The latency in milliseconds at high contrast was compared to that at medium contrast for each model cell. At each x-y point, the number of model cell results is indicated by color, where red corresponds to the highest density and dark blue to the lowest. The color is scaled as the log of the probability. A, Latency comparison when the threshold parameter for model cells was fixed across contrasts ($\theta=1.6$). B, Latency comparison when the threshold parameter at high contrast ($\theta=1.6$) was greater than that at medium contrast ($\theta=0.4$). C, Histograms of the probability of contrast-dependent latency differences. The dashed curve shows results from fixed-parameter data shown in (A), and the solid curve shows results from threshold-changing data shown in (B)

**Latency Variability within Individual Cells**

Above, we characterize the latency of cells by calculating a single number: the time of the first-peak of the spike-triggered average. However, the spike-triggered average can be confounded by a variety of factors, such as the number of spikes elicited in response to a stimulus. Therefore, we also compared the times of peaks in the time-varying firing rate across contrasts.

Peaks within the time-varying firing rates were identified as spiking events (see Methods). The times of spiking events can be conserved across contrasts, so a blind observer conservatively identified which events were "shared" across contrasts. Because this analysis can require more data than the calculation of a spike-triggered average, this analysis was performed for 8 of the ON cells and 8 of the OFF cells.

Our previous spike-triggered average calculations indicated that ON cells decreased their latencies of response at lower contrasts, opposite to the latency increase we observed in OFF cells and that others have reported in both cell types using other visual stimuli. This analysis similarly showed that the times of the spiking events at lower contrasts precede the times of the corresponding spiking events at higher contrasts (symbols are below the y=0 line in Figure 6).

We also found that the latency difference varied among the spiking events within individual cells (symbols are scattered along the y-axis in Figure 6). This variability could be at least partially explained by differences in the slope of the generator potential: the steeper the slope of the generator potential, the less an additive shift will affect the time of threshold crossing. Above we theorize that for ON cells the baseline generator potential is closer to threshold when contrast is lower. This would imply that that the higher-contrast events will substantially lag lower-contrast events when the slope of the generator potential preceding the event is shallow (for example, $t$=350-390 ms in Figures 3B, 3D and 3E). Little to no difference in the event times is predicted when the slope of the generator potential preceding the event is steep (for example, $t$=450-500 ms in Figures 3B, 3D and 3E).

To quantify this prediction, we calculated for each firing event the average slope of the most recent increasing segment of the generator potential preceding the event. As shown in Figure 6, when the slope of the generator potential was shallow (closer to 0), the higher-contrast events are substantially later than the corresponding lower-contrast events. When the slope of the generator potential was steep, there is little to no difference in the event times.

OFF cells also exhibited within-cell variability in the latency differences of spiking events. Interestingly, responses of most OFF cells included both spiking events for which high-contrast responses preceded lower-contrast responses and the converse (data not shown). Few cells showed a strong and significant correlation between generator potential slope and latency difference.

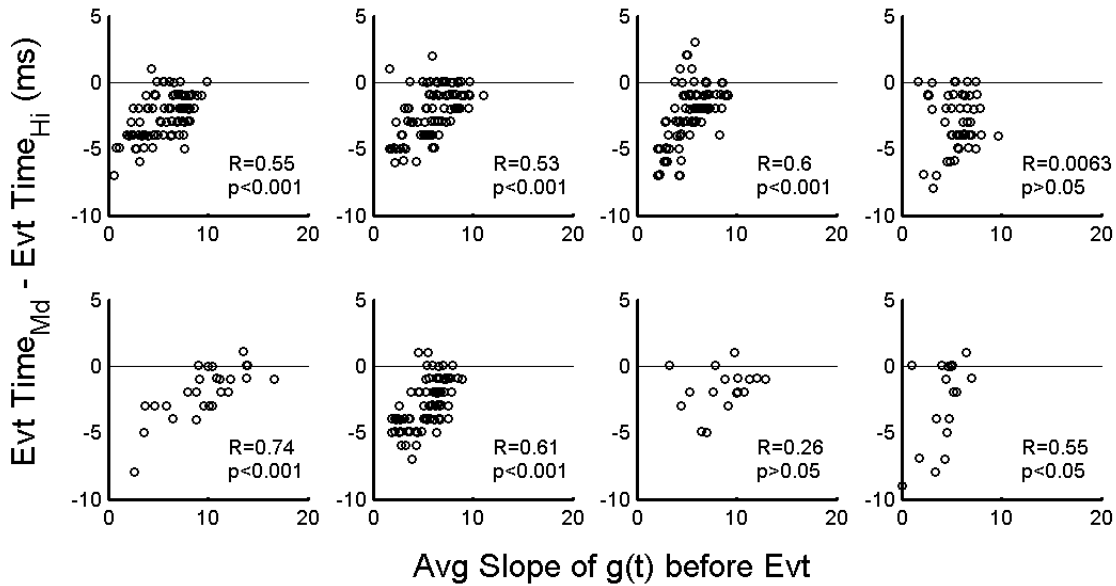

**Figure 6.** Latency differences for spiking events for ON cells depend on generator potential slope. Each panel corresponds to one ON cell. Each symbol corresponds to one spiking event identified as an event shared across contrasts. For each event, an event time difference was identified as the time of the event at medium contrast minus the time of the corresponding event at high contrast. (Values below the y=0 line indicate that the medium-contrast event preceded the high-contrast event). These values were compared to the average slope of an increasing segment of the generator potential, $g(t)$, preceding the event.

## *Discussion*

**Mechanistic Hypothesis to Explain Opposing Latency Effects**

Neurons of the early visual system have long been reported to exhibit longer-latency responses for low-contrast stimuli than for high-contrast stimuli. We report that LGN cells can under some conditions exhibit the opposite trend (Figures 1, 2 and 6). Both spike-triggered average analysis and spiking event-based analysis support this finding. We propose that at high contrast, an increase in inhibition causes an increase in the surround-to-center ratio, an increase in the distance-to-threshold, and therefore an increase in response latency.

Although contrast adaptation is more often reported, there is precedent for an additive mechanism in neurons of cat visual cortex (Carandini and Ferster, 1997). Within the LGN, the tonic hyperpolarizing current through $GABA_A$ receptors is one candidate for providing the postulated additive shift in membrane potential (Cope et al., 2005). No additive adaptation has been described in the retinal ganglion cell inputs to the LGN, but comparable data do not exist to exclude this possibility.

An increase in the strength of surround suppression could be attributed either to an increase in synaptic inhibition, or withdrawal of synaptic excitation. Here we consider in more detail the possible role of synaptic inhibition at the level of the LGN. LGN neurons receive direct or indirect inhibition through local feed-forward inhibitory pathway, a loop through the reticular nucleus, and a feedback loop through visual cortex. Feed forward inhibition in the LGN is both powerful and rapid. Inhibitory synaptic input to the receptive field center has been shown to underlie other temporal computations underlying the visual control of bursting (Wang et al 2007). In the case of contrast adaptation, we suggest that changes in temporal contrast result in changes in the spatial opponency of the receptive field, specifically, an increase in the inhibitory input to the opponent surround region. This hypothesis makes the experimental prediction that the adaptation of spatial antagonism should be stronger in the ON pathway than the OFF pathway, though no such asymmetry has yet been reported.

Blitz and Regehr (2005) reported that many LGN relay neurons receive nonlocked inhibition from LGN interneurons, and suggest that such inhibition may account for receptive field surround inhibition properties. The nonlocked inhibition received by a specific LGN cell was elicited by activation of retinal ganglion cells that do not synapse directly onto the cell. This inhibition most likely reflects disynaptic inhibition from retinal ganglion cells with the opposite light sensitivity to the cell's excitatory inputs (the "pull" of the push-pull receptive field). One possibility is that at high contrast, LGN relay cells receive nonlocked inhibitory input from a wider spatial field of LGN interneurons, increasing the suppressive surround of the LGN receptive field. It is not clear, however, why such a mechanism would have an additive rather than multiplicative effect.

Although there is little evidence for ON-OFF asymmetries in LGN anatomy, the effect of feed-forward inhibition can be substantially different depending on whether the LGN neuron is in a burst mode or a tonic spiking mode (Blitz and Regehr, 2005). In Figure 2F-G and the associated text, we report that the probability of burst responses increased to above 20% for three ON cells. The latency at low contrast for these three ON cells was greater than the latency at higher contrasts, in contrast to the general latency trend observed for ON cells. The difference in the effect of feed-forward inhibition may at least partially account for the differences in these cells' latency trends.

An alternative abstract model of the LGN has been presented that accounts for a wide range of suppressive effects in terms of contrast gain control at the level of neural firing rate (Bonin et al, 2005). That model is not designed to make predictions regarding spike latency, and does not incorporate an additive component of contrast adaptation, and thus we believe it addresses a distinct component of contrast adaptation. It is unclear which inhibitory inputs contribute the multiplicative suppression described by Bonon et al, and which underlie the additive effect we describe.

**Previously Reported Latency Trends**

The mechanism we hypothesize above can account for our data as well as apparently disparate previous reports. First, the majority of the previous studies that reported a consistent latency decrease with higher contrast were studies of retinal neurons (eg., Baccus and Meister, 2002; Benardete and Kaplan, 1999; Chander and Chichilnisky, 2001; Kaplan and Benardete, 2001; Kim and Rieke, 2001; Shapley and Victor, 1978; Victor, 1987; Zaghloul et al., 2005). Therefore, LGN interneurons would not yet be able to influence the responses.

Alternatively or in addition, the stimulus used in many previous studies differed from that of the current experiment. Full-field stimuli may cause more nonlocked inhibition to an LGN relay neuron: the LGN relay neuron may receive nonlocked inhibition from a greater number of interneurons and/or the LGN relay neuron may receive more nonlocked inhibition from a single interneuron if many ganglion cells synapse onto the interneuron. Further, many previous studies used Gaussian stimuli or sum-of-sinusoid stimuli, whereas the current study used binary stimuli. Our high-contrast stimulus may have thus been of substantially higher contrast than other studies' high-contrast stimulus.

To determine the relative importance of the cell type and the stimulus, it will be of critical importance to learn whether retinal ganglion cell responses also show latency advance under our stimulus conditions.

**ON-OFF Cell Asymmetry**

We report two ON-OFF cell asymmetries: a contrast-dependent latency asymmetry (Figure 2) and a contrast-dependent shift in the nonlinear spiking function (Figure 4). Both could be explained by a contrast-dependent change in distance to threshold in ON cells. We cannot at present specifically explain why ON and OFF cells show these asymmetries, although we consider that our reported asymmetries may relate to other previously reported differences between the pathways. We do not know if the opposing latency effect arises in the LGN, or whether this ON-OFF asymmetry may trace back to the retina.

Although the average latency change for OFF cells was generally in the expected direction, individual spiking events of OFF cells could exhibit an opposite latency change to the general trend (Figure 6). If the additive shift is attributable to inhibitory input at the level of the LGN, it is possible that both the ON and OFF LGN relay neurons receive nonlocked inhibitory input from LGN interneurons but that the inhibitory input is more sustained for the ON pathway. We hypothesize that ON retinal ganglion cells provide more excitation to LGN interneurons than OFF retinal ganglion cells, thereby causing ON LGN relay cells to receive more sustained inhibition. Zaghloul et al. (2003) reported that the maintained firing rate for ON retinal ganglion cells (18.4 ± 3.8 Hz) is larger than that for OFF retinal ganglion cells (6.6 ± 1.3 Hz). Therefore, the ON retinal ganglion cells may excite a given LGN interneuron more than do OFF retinal ganglion cells. Additionally, ON retinal ganglion cells have been reported to have larger receptive fields than OFF retinal ganglion cells (Chichilnisky and Kalmar, 2002). We consider the possibility that in the ON pathway, more retinal ganglion cells synapse onto an LGN interneuron, and/or that more LGN interneurons synapse onto an LGN relay cell.

Blitz and Regehr (2005) reported that only a subset of LGN relay cells received nonlocked inhibition, and Lesica et al. (2007) reported that only a subset of the LGN cells showed an increase in the surround-to-center ratio with contrast. We would predict that more ON cells are included in these subsets than OFF cells.

**Low-Contrast Specific Spiking Events Predictions**

We have forwarded the hypothesis that the distance to threshold can change with contrast for LGN neurons. We note that this hypothesis makes the further prediction of low-contrast specific firing events. If a change in the contrast merely scaled the generator potential without producing a change in the distance from the baseline generator potential to the threshold, all spiking events (identified by peaks in the time-varying firing rate) at low threshold should have corresponding spiking events at high contrast. For example, in Figure 3A every time the convolved low-contrast stimulus crosses the threshold, there is a corresponding instance that the convolved high-contrast stimulus crosses the threshold. This is because every time the convolved low-contrast stimulus crosses the threshold, the corresponding convolved high-contrast stimulus is larger (as it is merely a scaled version of the convolved low-contrast stimulus). Therefore, it necessarily also crosses threshold.

On the other hand, if the distance-to-threshold is reduced at lower contrasts, low-contrast specific spiking events are predicted (eg., Figure 3D shows blue peaks without corresponding red peaks at approximately 80 and 445 ms). Consistent with our hypothesis, in LGN neurons we find that while many spiking events are conserved across contrasts, some spiking events are only present during low contrasts (see Figure 3E). The times of these low-contrast specific spiking events are the same as those predicted by the "depolarized" convolved low-contrast stimulus (compare Figure 3D and 3E).

**Additive Shift versus Multiplicative Gain Change**

Figure 4E-G suggests that in the LGN, the magnitude of the contrast-dependent change of the distance-to-threshold (the shift in the nonlinear function) varies across cells. Interestingly, we also find that cells exhibiting strong multiplicative gain changes are less likely to show a horizontal shift in their non-linear functions (see Figure 4H).

Conceptually, both the gain changes and the additive shift may address the same problem: encoding a range of stimuli using a fixed distribution of firing rates. Encoding a range of contrasts presents the difficulty of avoiding firing rate saturation at high contrasts while also preserving sensitivity to small fluctuations at low contrasts. Multiplicative gain change indicates that a neuron is more sensitive to changes in the stimulus at low contrasts than at high (eg., a scaling of the generator potential). The additive shift indicates that all stimuli are closer to threshold, thereby improving the probability of a resultant spike (eg., the addition of a constant to the generator potential). Thus, while neurons show variability both the extent of gain change and additive shift, either of these effects or a combination thereof may serve the common purpose of improving encoding across a range of contrast conditions. Cell-to-cell variability in the extent of additive shift versus multiplicative gain change may relate to bipolar cells contributing to the response. Baccus and Meister (2002) reported that bipolar cells showed either an additive or a multiplicative gain change, but not both.

**Conclusion**

Under specific stimulus conditions, some LGN neurons can exhibit shorter latency responses at low contrast than at high contrasts. Physiological and model data supports the hypothesis that the distance-to-threshold is lower at low contrasts in those cells, which may account for the unexpected direction of the latency change. We hypothesize that ON LGN relay cells receive more inhibitory input at high contrasts, thereby producing a larger surround-to-center ratio, resulting in an additive form of contrast adaptation that can be revealed by longer response latency.

## *Experimental procedures*

**Surgical preparation.** Cats were initially anesthetized with ketamine HCl (20 mg/kg, i.m.), followed by sodium pentothal (2-4 mg $*$ kg$^{-1}$ $*$ h$^{-1}$, i.v., supplemented as needed). Cats were ventilated using an endotracheal tube. Electrocardiogram, electroencephalogram, temperature, expired $CO_2$, and oxygen in blood were continuously monitored. Surgical and experimental procedures were in accordance with National Institutes of Health and United States Department of Agriculture guidelines and were approved by the UCSD Institutional Animal Care and Use Committee.

**Electrophysiology.** We recorded from 41 LGN relay cells, sampled the four main cell classes: ON X, OFF X, ON Y, and OFF Y. Parylene-coated tungsten electrodes (AM Systems, Everett, WA) were inserted through a 0.5 cm diameter craniotomy over the LGN. Electrical recordings were amplified, filtered, and digitized at 10kHz sampling rate (CED micro 1401 and Spike2, ver. 5.12a; Cambridge Electronic Design, Cambridge, UK). Waveforms were analyzed offline isolate single unit responses (Fee et al., 1996).

**Visual stimulation.** Stimuli were full-field and presented on a custom-built LED array. We began with a random binary stimulus of 125 frames/s for 10 s. The same binary stimulus was scaled about the mean to obtain three contrast conditions (11%, 33%, and 100%, where contrast is defined as the standard deviation of the luminance over the mean). Stimuli of the three contrasts were interleaved and presented between 10 and 128 times each. We only analyzed the last five seconds of the response to each 10-s stimulus. The mean luminance was constant across contrasts and was within the photopic range.

**Response Latency**. Spike-triggered averages (STAs) were computed for each cell at each contrast. The latency was defined as the time to the shortest latency peak of the STA. Because of noise in our STA, we estimated the time of the peak from a Gaussian fit as follows: we define $t_P$ as the time at which the absolute value of the STA reached its maximum between -40 and 0 milliseconds before the spike. We determined the nearest time before and after $t_P$ ($t_A$ and $t_B$) at which the STA crossed the mean stimulus value. The STA between $t_A$ and $t_B$ was fit to a Gaussian. The latency was defined as the time of the peak of the Gaussian. We repeated latency analysis by comparing the times to the second peak of the STA across contrasts. Results were unchanged.

For 16 of the LGN cells (8 ON cells and 8 OFF cells), we also analyzed the latencies of individual spiking events. For each cell, at each contrast, we determined the time of the firing events (PSTH peaks). The times of the events were identified as in Berry et al., (1997), but we determined the event time as that for which the smoothed PSTH reached its first maximum within the event boundaries. A blind observer conservatively identified which of these events were shared across contrasts.

**Linear-Nonlinear Cascade.** For each cell at each contrast, we first estimated the filter as the spike-triggered average. The filters for all contrasts were normalized by the amplitude of their first peaks. We convolved the stimulus by the corresponding filter to create a generator potential, $g(t)$ and then compared this generator potential to the observed probability of spiking at each time bin. For each cell, we fit this observed input-output function at 100% contrast to the following sigmoid equation:

$$y = \exp(A - e^{-Gx+S}) \qquad \text{Eq. 1}$$

where $A$, $G$, and $S$ describe the amplitude, slope, and horizontal offset of the nonlinear function, respectively. We then fit the input-output functions of the other two contrasts, holding the amplitude ($A$) constant for all contrast conditions for each cell. Data sets were excluded from our analysis if there were less than 100 spikes observed in the response, or if the $R^2$ value associated with the sigmoid fit was less than 0.90. For each cell at each contrast, we then define the Gain as $G$ and the Offset as $S$ from Eq. 1. Our results did not depend critically on how the filters were scaled; similar results were found by normalizing by the peak-to-peak amplitude of the spike-triggered average.

In Figure 4I, we use the linear-nonlinear cascade model to predict the time-varying firing rate of an LGN neuron during the low-contrast stimulus condition. As described above, we first calculated the linear filter and the Gain G and Offset S of the input-output function for all three contrast conditions. We then convolved the low-contrast stimulus with the low-contrast linear filter. We defined three input-output functions, each of which used the Gain G from the low-contrast condition, but differed in that the Offset S was either from the high, medium or low-contrast condition. Based on the convolved stimulus at each time point, we could then estimate the probability of spiking at a function of time using each of these input-output functions.

**Contrast Normalization Index.** The greater the contrast change, the larger gain (sensitivity) change would be required to compensate. Therefore, we express the magnitude of gain change relative to the magnitude of contrast change and call this the contrast normalization index $\kappa$:

$$\kappa = \frac{G_{lower}/G_{higher} - 1}{C_{higher}/C_{lower} - 1} \quad \text{Eq. 2}$$

where *G* is neural gain and *C* is stimulus contrast.

**LGN Spiking Model.** Model cells were implemented as described in Keat et al., (2001). Briefly, the model first convolves the stimulus with a linear filter. The generator potential is equal to the convolved stimulus plus noise; the amplitude and time-constant of this noise are defined by the model parameters $\sigma_a$ and $\tau_A$, respectively. A spike is generated whenever the generator signal crosses a threshold, $\theta$. Each time a spike occurs, a negative after-potential is added to the generator potential, such that the threshold is crossed repeatedly during sustained excitatory stimuli. The amplitude, time-constant, and variability in the amplitude of the negative after-potential are defined by the model parameters B, $\tau_P$, and $\sigma_b$, respectively.

With regard to the analysis presented in Figure 5, we generated a set of model cells using random combinations of the following first parameters: B = 3, 5, or 7; $\tau_P$ = 20, 35, or 50; $\tau_A$ = 20; $\sigma_a$ = 0.01, 0.16, 0.31, or 0.61, $\sigma_b$ = 0.02, 0.15, or 0.28. The filter function, F, was defined as $F = 2\sigma_1^2\pi \cdot \exp(-(x-\mu_1)^2/(2 \cdot \sigma_1^2)) - \sigma_2^2\pi \cdot \exp(-(x-\mu_2)^2/(2 \cdot \sigma_2^2))$. The parameters for F, were randomly chosen combinations from the following second parameter ranges: $\mu_1$ = [-45,-20]; $\sigma_1$ = [10,18]; $\mu_2$ = [-70,-35]; $\sigma_2$ = [18,27], with the restriction that F(x=0) be less than or equal to 5% of the maximum value of F. After the first and second parameters were chosen, data was generated for each contrast condition for two thresholds: $\theta$ = 0.4 and 1.6.

**Acknowledgements**

This work was supported by NEI R01-EY016856; KG was supported by an NSF graduate research fellowship.